\documentclass[twocolumn,letterpaper,amsmath,amssymb,floatfix,prl,superscriptaddress]{revtex4} 

\usepackage{graphicx}
\usepackage{dcolumn}  
\usepackage{bm}  
\usepackage{epsfig}

\begin{document}
\setlength\arraycolsep{2pt} 

\title{Overscreening in 1D lattice Coulomb gas model of ionic liquids}

\author{Vincent D\' emery}
\affiliation{Laboratoire de Physique Th\' eorique (IRSAMC), Universit\' e de Toulouse, UPS and CNRS, F-31062 Toulouse, France}

\author{David S. Dean}
\affiliation{Laboratoire de Physique Th\' eorique (IRSAMC), Universit\' e de Toulouse, UPS and CNRS, F-31062 Toulouse, France}

\author{Thomas C. Hammant}
\affiliation{DAMTP CMS, Wilberforce Road, Cambridge CB3 0WA, United Kingdom}

\author{Ronald R. Horgan}
\affiliation{DAMTP CMS, Wilberforce Road, Cambridge CB3 0WA, United Kingdom}

\author{Rudolf Podgornik}
\affiliation{Department of Theoretical Physics, J. Stefan Institute and Department of Physics, Faculty of Mathematics and Physics, University of Ljubljana, SI-1000 Ljubljana, Slovenia}

\begin{abstract}
Overscreening in the charge distribution of ionic liquids at electrified interfaces is shown to proceed from purely electrostatic and steric interactions in an exactly soluble
one dimensional lattice Coulomb gas model. Being not a mean-field effect, our results suggest that even in higher dimensional systems the overscreening could be accounted for by 
a more accurate treatment of the basic lattice Coulomb gas model, that goes beyond the mean field level 
of approximation, without any additional interactions.  
\end{abstract}

\maketitle

Room temperature ionic liquids (RTILs) are Coulomb fluids with large, asymmetric ions, and
are used as electrolytes in fuel and solar cells, batteries and supercapacitors, to name but a few  important applications \cite{review1}. As pointed out in several seminal contributions by  Kornyshev \cite{Kornyshev1}, the size of the ionic species leads in general to crowding and lattice saturation,  thus engendering a fundamentally different behavior of  ionic liquids  at charged interfaces as compared to  aqueous electrolytes. In some particular cases several aspects of this behavior can be captured on a mean-field level by the lattice Coulomb gas (LCG) Kornyshev model \cite{Bazant-rev,Kornyshev1}.  

Steric effects stipulate that the capacitance of ionic liquids decays at large applied voltages while
at the point of zero charge (PZC) it can exhibit a maximum as well as a minimum (and is thus 
a nonmonotonic  function of applied voltage) depending on the lattice packing fraction \cite{Kornyshev2}, whereas 
for dilute electrolytes the capacitance at PZC is always a minimum \cite{Kornyshev1}. 
X-ray reflectivity \cite{Metzger}, SFA \cite{SFA} and AFM \cite{AFM} studies at charged interfaces reveal an alternating charge distribution starting with an {\sl overscreening} cationic layer, at the negatively charged substrate, which decays roughly exponentially into the bulk liquid with a periodicity comparable with the size of ionic species \cite{Perkin}.  This observed charge  layering is expected to be a generic feature of RTILs at charged interfaces resulting from an interplay between steric effects and strong electrostatic coupling. While the dependence of the differential capacity can be predicted within the mean-field  solution of the LCG model \cite{Kornyshev2}, the overscreening and alternate charge layering at an electrified interface can not.  

In order to explain overcharging and charge oscillations, Bazant {\sl et al.}  \cite{Bazant} proposed a phenomenological theory based on a Landau-Ginzburg-like functional containing the standard LCG free energy \cite{Borukhov} but with an additional higher order potential-gradient term, similar to what is found in Cahn-Hilliard models, and then solving it on a mean-field level. However, this higher order gradient term can also be seen \cite{Santangelo} as stemming from a decomposition of the Coulomb interaction into a long-distance mean-field-like component and a non-mean-field strong coupling component \cite{Kanduc}. This alternative interpretation motivates a more detailed non-mean-field analysis of the original Kornyshev LCG model in order to go beyond the limitations of the mean-field approximation. We thus propose an exact  analysis, albeit in one dimension, that demonstrates the full physical phenomenology  of the LCG model, and in particular shows that overscreening emerges naturally from this model without the introduction of any new physical interactions.

We solve exactly  the statistical mechanics of a one dimensional LCG in a standard condenser configuration, with a method that generalizes other treatments of the point-like, or continuous, one dimensional Coulomb gas models extensively used in the study of electrolytes \cite{oned}. We consider a system where charges $q$ (cations) or $-q$ (anions) are located on a line at lattice points with lattice spacing $a$ and a total size of $M$ points. The Hamiltonian due to Coulomb interactions is
\begin{equation}
\beta{\cal H}[S_i] = - \frac{\gamma}{4}\sum_{i,j = 0}^{M-1} \vert i - j\vert S_iS_j
\end{equation}
where $\gamma = \frac{\beta q^2 a}{\epsilon\epsilon_0}$ is the ratio of the electrostatic to thermal energy. Here $S_i$ is a classical spin variable taking the value $S_i = 1$ if there is a cation at lattice site $i$, $S_i = -1$, if there is an anion, and $S_i = 0$ if the site is unoccupied. We will impose overall electroneutrality on the system, stipulating that $\sum_i S_i = 0$. The grand canonical partition function can then be written as
\begin{equation}
\Xi = {\rm Tr}_{S_i} \left[ \mu^{\sum_{i=0}^{M-1} \vert S_i\vert} \int_{-\pi}^{+\pi} \!\!\!\!\!\!\exp\left( \beta{\cal H}[S_i]  + i \psi \sum_{i=0}^{M-1} S_i \right) \frac{d\psi}{2\pi}\right],
\end{equation}
where $\mu$ is the fugacity of  both anions and cations, assuming for simplicity  that the ionic liquid is symmetric. Carrying out  a Hubbard-Stratonovich transformation, the grand canonical partition function can be expressed  as a path integral over a field $\phi_j$ on the lattice, while the integral over $\psi$ corresponds to the integral over $\phi_0$, but runs over an interval of length $2\pi$. The partition function then assumes the form
\begin{equation}\label{xi}
\Xi=\int \prod_{j=0}^{M-1} \frac{d\phi_j}{\sqrt{2\pi\gamma}}\exp\left(-{\cal S}[\phi] \right).
\end{equation}
where
\begin{equation}
{\cal S}[\phi] =\sum_{j=0}^{M-2} \frac{(\phi_{j+1}\!-\!\phi_j)^2}{2\gamma} - \sum_{j=0}^{M-1} \log(1+2\mu\cos(\phi_j)).
\label{action}
\end{equation}
In the case where there are  external charges on the boundaries of the system, $+qQ$ on the site $-1$ and $-qQ$ on the site $M$ in codenser configuration, we need to add $iQ(\phi_{-1}-\phi_M)$ to the action (\ref{action}). The field $\phi$ can be identified with
the fluctuating electrostatic potential {\sl via} the relation $V= -i\phi/\beta q$. One can  show that the saddle-point equation for  the above functional integral, $\frac{\delta {\cal S}}{\delta \phi_j}\vert_{\phi_\textrm{MF}} = 0$,  reduces to the mean-field equations of  Kornyshev \cite{Kornyshev1} and Borukhov {\sl et al.} \cite{Borukhov}, if one takes the continuous limit ($a \rightarrow 0$ with the maximal charge density $q/a$ kept constant).

Writing $y_i=\phi_i$ and defining  $p^{1/2}(y,y') = \frac{1}{\sqrt{\pi\gamma}} \exp\left(-\frac{(y-y')^2}{\gamma}\right)$ we introduce the symmetric operator 
\begin{equation}\label{xi1}
K(y,y')=\int p^{1/2}(y,z)(1+2\mu\cos(z))p^{1/2}(z,y')dz,
\end{equation}
which allows us to write the grand potential $\Omega_Q$  at fixed  external charges  $\pm Q$ as 
\begin{equation} 
\exp({-\beta \Omega_Q})= \int_{-\pi}^\pi\!\!\!\!dx\int_{-\infty}^\infty\!\!\!\!\!\!dy \  e^{iQx}\!\!\left[p^{1/2}K^M p^{1/2}\right]\!\!(x,y)e^{-iQy}.\label{xiq}
\end{equation}
Introducing the ket-vector $|\psi_Q\rangle=p^{1/2}|e^{-iQy}\rangle$, we can write
\begin{equation}
\exp({-\beta \Omega_Q}) =\langle\psi_Q|K^M|\psi_Q\rangle .
\end{equation}

If instead of fixing the  charge at the boundaries one keeps fixed the potential difference between them, $\Delta V = \Delta \nu/\beta q$, 
it is easy to show that
\begin{equation}\label{xiv}
 \exp(-\beta \Omega_{\Delta \nu}) = \int d\sigma \exp(- \Delta \nu \sigma-\beta \Omega_\sigma)
\end{equation}
where $\Omega_{\Delta \nu}$ is now the grand potential at imposed external potential difference. It follows  that the average charge and differential capacity both as a function of the imposed potential difference are given by 
\begin{equation}
\langle Q\rangle_{\Delta \nu}=-\frac{\partial (\beta \Omega_{\Delta \nu})}{\partial \Delta \nu} \qquad {\rm and} \qquad c_{\Delta \nu}=\frac{\partial \langle Q\rangle_{\Delta \nu}}{ \partial \Delta \nu}.
\end{equation}

To obtain the charge density we need to know the average charge number $\langle S_i\rangle$ on site $i$. Therefore we just need to replace $1+2\mu\cos(y_i)$ with $i ~2\mu \sin(y_i)$ in Eq. (\ref{xi1}), via a new operator
\begin{equation}
L(y,y') = i \int_{-\infty}^\infty p^{1/2}(y,z)2\mu\sin(z)p^{1/2}(z,y')dz.
\end{equation}
Putting charges $Q$ and $-Q$ on the left and right boundaries, the mean charge density on site $i$ is then
\begin{equation}
\langle S_i\rangle=\frac{\langle\psi_Q|K^iLK^{M-i-1}|\psi_Q\rangle}{\langle\psi_Q|K^M|\psi_Q\rangle} .
\end{equation}
In order to compute the thermodynamic quantities derived here, we need to evaluate 
the indicated matrix elements numerically. The domain of definition of the operators $K$ and $L$ depends on the 
choice of the surface charge $Q$. If we define the $Q$-dependent class of functions 
\begin{equation}\label{clfunc}
f(x)= \sum_{k\in\mathbb{Z}-Q} {\tilde f}_k \exp(ikx),
\end{equation}  
they are stable under the action of $K$:
\begin{eqnarray}
\widetilde{Kf}_k &= &e^{-\gamma k^2/4}\left[ e^{-\gamma k^2/4}\tilde f_k \right. \\
& &+\mu \left.\left(e^{-\gamma (k+1)^2/4}\tilde f_{k+1}  +e^{-\gamma (k-1)^2/4}\tilde f_{k-1}\right)\right]. \nonumber
\end{eqnarray}
The action of the operator $K$ and $L$, can thus be carried out numerically in this Fourier representation.  Numerical computations  are performed by truncating the Fourier components at a maximum wave vector  $k_\textrm{max}$. This approximation is valid if $\gamma k_\textrm{max}^2 \gg 1$ which means that in the cases studied here  we can take  $k_\textrm{max} = 25$.

Numerically one can compute  (i)  the grand potential of the system and the disjoining pressure giving the effective interaction between the two bounding layers (ii) the differential capacitance as a function of the voltage difference between the boundaries and (iii) the dependence of the charge density on the lattice position.

\begin{figure}[t!]
\centerline{\psfig{figure=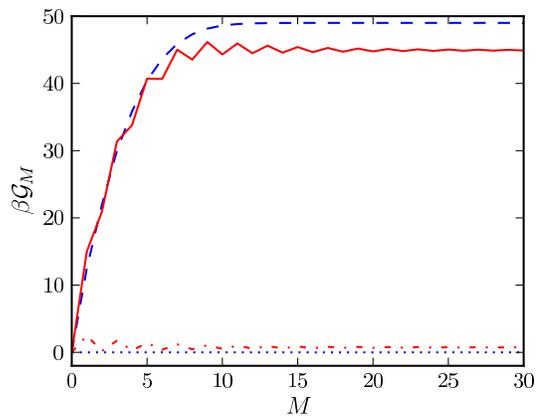,width=0.85\linewidth}}
\caption{Dimensionless free enthalpy $\beta {\cal G}_M$ as a function of the system size $M$, for the fugacity $\mu=100$: solid line: exact, $Q=5$; dashed line: mean field, $Q=5$; dash-dotted line: exact, $Q=0$; dotted line: mean field, $Q=0$. }
\label{f_M_J}
\end{figure}

First, we look at the grand potential $\Omega$ as a function of the size of the system $M$ for given charges $\pm Q$ on the boundaries. The discrete pressure on the boundaries is $P={\Omega}_M-{\Omega}_{M+1}$ and tends to the bulk pressure $P_b$ when $M$ is large. The actual force acting on the boundaries of the system is given by the disjoining pressure $P_d=P-P_b$ that follows from the free enthalpy $\mathcal{G}_M={\Omega}_M + M P_b$, which is plotted on Fig \ref{f_M_J}. at high density and for different values of $Q$. The mean field results are also shown for comparison.

We observe two salient features of the disjoining pressure: (i) it is attractive when the system size is small and goes to zero for $M \gtrapprox 2Q$ (ii) it exhibits oscillations of positive/negative pressure when $M$ is even/odd. When the size of the system is large, the boundary charges are completely screened by the ions, and the plates only feel the bulk pressure; when the boundary layers are close to each other, not enough space is available for ions to screen the boundary charges and they thus attract each other because of their opposite charge. Having an even number of sites obviously stabilizes the system compared to an odd number, however, this  effect vanishes as the size increases. This behavior is not seen within the mean-field approximation and is due to the fact that an odd number of sites  cannot all be filled due to the electroneutrality condition, and this induces a high energy cost when $\mu$ is large. When the system becomes large this effect is asymptotically negligible, and it does not show up at lower densities.  The mean field result does not show these oscillations, but the general trend agrees with the exact result as well as with the continuous model mean-field results \cite{Ben-Y}. Another difference that is not shown here is that the bulk pressure is lower in the exact result than in the mean-field result.

The bulk pressure too has some subtle properties. It is given by $P_b=\ln(\lambda_0)$, where $\lambda_0$ is the highest eigenvalue of $K$. But the operator $K$ acts on the set of functions (\ref{clfunc}), which depends on the non integer part of $Q$, $\theta=Q-\lfloor Q\rfloor\in [0,1)$. Thus we may expect that the bulk pressure depends on $\theta$. This is indeed the case, but this dependence is very small: $\Delta P_b/P_b\sim 10^{-3}$. This dependence corresponds to the so-called {\sl $\theta$-vacuum} introduced in Ref. \cite{Aizen} for a standard Coulomb gas.

\begin{figure}[t!]
\begin{center}
\includegraphics[angle=0,width=0.85\linewidth]{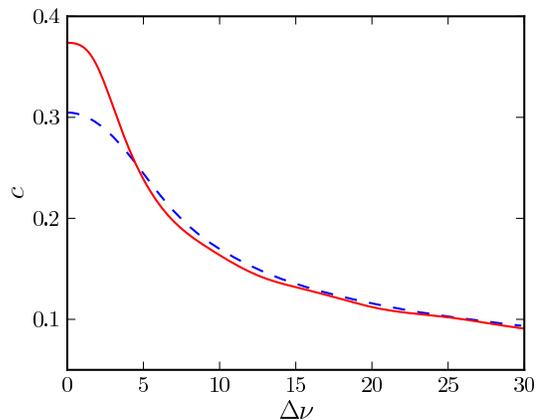}
\end{center}
\caption{Capacitance as a function of the voltage drop for $\mu=10$ and $\gamma=1$: exact result (solid line) and mean field result (dashed line).}
\label{f_V_C_bell}
\end{figure}

\begin{figure}[t!]
 \begin{center}
\includegraphics[angle=0,width=0.85\linewidth]{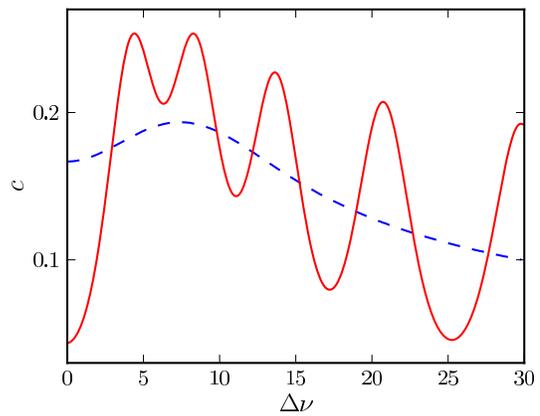}
\end{center}
\caption{Capacitance as a function of the voltage drop for $\mu=0.1$ and $\gamma=1$: exact result (solid line) and mean field result (dashed line). For smaller $\gamma$ the non-monotonicy gradually disappears and the exact solution approaches the mean-field result of Kornyshev \cite{Kornyshev1}.}
\label{f_V_C_camel}
\end{figure}

We next analyze the differential capacitance as a function of the boundary voltage difference for $M=100$ sites and $\gamma=1$ at high density ($\mu=10$) on Fig. \ref{f_V_C_bell} and low density ($\mu=0.1$) on Fig. \ref{f_V_C_camel}. Two phenomena emerge in the behavior of capacitance: (i) the capacitance has a dip at PZC (ii) the capacitance exhibits oscillations, distinctly visible at low densities. The PZC dip appears at low $\mu$ and confirms the continuum mean-field results of Kornyshev \cite{Kornyshev1}. The oscillations stem from the $\theta$-dependence of the bulk pressure inducing an extensive $\theta$-dependence of the grand potential, which then exhibits minima at $\theta^*$ corresponding to boundary charges $Q\in\mathbb{Z}+\theta^*$. When the imposed voltage varies, the average charge exhibits plateaus at these selected charges, and the jumps between plateaus induce the peaks in the capacitance. These peaks become weaker as the system gets smaller or as the temperature increases. 

The mean-field capacitance exhibits exactly the same trends but without oscillations,  which is consistent with the absence of plateaus in the average charge. Again our mean-field results coincide with those of ref. \cite{Kornyshev1}.

\begin{figure}
 \begin{center}
\includegraphics[angle=0,width=0.85\linewidth]{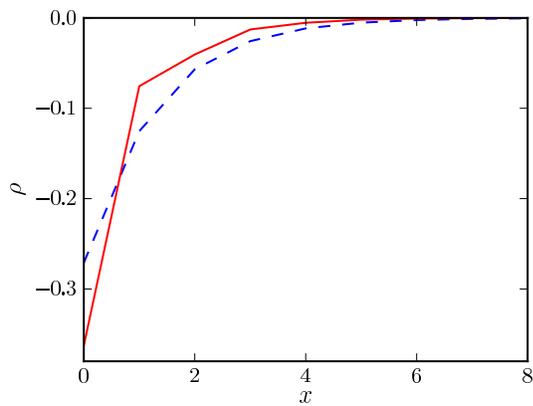}
\end{center}
\caption{Mean charge density close to the left electrode (located at $x=-1$) as a function of the position for $Q=0.5$, $\gamma=1$ and $\mu=1$: exact result (solid line) and mean field result (dashed line).}
\label{f_x_rho}
\end{figure}

Next we analyze the charge density profile in the vicinity of one of the boundaries (left); the same profile with opposite charge is obviously found close to the other boundary. First we note that for small packing fractions ($\mu=1$) the charge density shows a monotonic variation as a function of the separation from the boundary, Fig. \ref{f_x_rho}, and agrees  with the mean-field result \cite{Borukhov}.

\begin{figure}
 \begin{center}
\includegraphics[angle=0,width=0.85\linewidth]{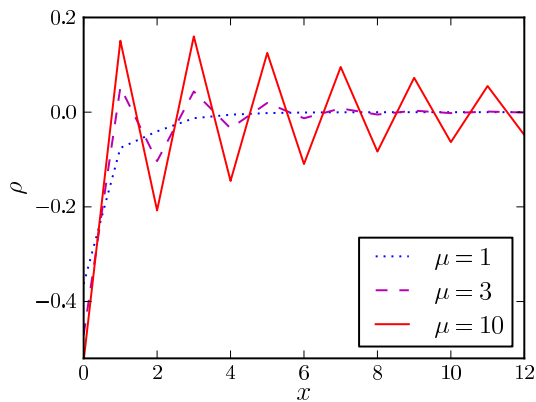}
\end{center}
\caption{Exact result for the mean charge density close to the left electrode (located at $x=-1$) as a function of the position for $Q=0.5$, for $\gamma=1$ and different values of the fugacity.}
\label{f_x_rho_layers}
\end{figure}

As we increase the packing fraction ($\mu>1$, Fig. \ref{f_x_rho_layers}) the overscreening effect starts to dominate the behavior of the charge density and clear charge layering emerges: a counterion layer is followed by a coion layer with their thicknesses equal to the lattice unit. This corresponds closely to the experimentally observed  situation \cite{Metzger} and approximately to the results of the model of Bazant {\sl et al.} \cite{Bazant} if the packing fraction is not  too large.

The charge distribution exhibits the following phenomenology: (i) the amplitude of the oscillations varies continuously with the boundary charge $Q$, being maximal for half integer values and zero for integer charges, due to the fact that in 1D one can have perfect screening only for integer values of $Q$ (ii) the overcharging oscillations are damped with a characteristic length $\xi$, with $\xi\sim\mu$.  The characteristic lengthscale of the overscreening oscillations can be obtained from the two highest eigenvalues of the transfer matrix $K$, $\lambda_0$ and $\lambda_1$, and the definition of the correlation length $\xi=\left(\ln \left|\frac{\lambda_0}{\lambda_1}\right|\right)^{-1}$. Analytic computations of these eigenvalues for $k_\textrm{max}=1$ and $Q=0$, and for $k_\textrm{max}=0.5$ and $Q=0.5$ indeed gives $\xi\sim\mu$. 

Since the mean-field results  show no overscreening, it must be a specific feature of the exact result for the same LCG model. We thus conclude that in order to describe the overscreening phenomena in ionic liquids one needs to go beyond the mean-field level of approximation. There appears to be no need to modify the original LCG Hamiltonian but it is essential to retain its discretized lattice form as opposed to its continuum limit. Based on our results, valid in 1D, we conclude that an approximation akin to the strong-coupling limit, as defined for ordinary Coulomb fluids \cite{review}, could be pertinent. However, deriving such an approximation systematically for a lattice Coulomb gas, where there are multiple length scales, is an open problem.

D.S.D. acknowledges support from the Institut Universitaire de France. R.P. acknowledges support of The Leverhulme Trust and illuminating discussions with S. Perkin, D. Harries and U. Raviv.

\end{document}